\documentclass[conference, anonymous]{IEEEtran}
\IEEEoverridecommandlockouts
\usepackage{cite}
\usepackage{amsmath,amssymb,amsfonts}
\usepackage{graphicx}
\usepackage{textcomp}
\usepackage{comment}
\usepackage{booktabs}
\usepackage{listings}
\usepackage{xcolor}
\usepackage{algorithm}
\usepackage{algpseudocode}
\usepackage{color}
\usepackage[most]{tcolorbox}
\usepackage{multicol}
\usepackage{stfloats}

\usepackage{lscape}
\usepackage{adjustbox} 
\usepackage{caption}
\usepackage{subcaption}

\usepackage{xcolor}
\def\BibTeX{{\rm B\kern-.05em{\sc i\kern-.025em b}\kern-.08em
    T\kern-.1667em\lower.7ex\hbox{E}\kern-.125emX}}

\usepackage{listings}
\usepackage{subcaption} 
\usepackage{xcolor}
\usepackage{enumitem}

\definecolor{codegreen}{rgb}{0,0.6,0}
\definecolor{codegray}{rgb}{0.5,0.5,0.5}
\definecolor{codepurple}{rgb}{0.58,0,0.82}
\definecolor{backcolour}{rgb}{0.95,0.95,0.92}

\lstdefinestyle{mystyle}{
    backgroundcolor=\color{white},   
    commentstyle=\color{codegreen},
    keywordstyle=\color{magenta},
    numberstyle=\tiny\color{codegray},
    stringstyle=\color{codepurple},
    basicstyle=\ttfamily\footnotesize,
    breakatwhitespace=false,         
    breaklines=true,                 
    captionpos=b,                    
    keepspaces=true,                 
    numbers=left,                    
    numbersep=5pt,                  
    showspaces=false,                
    showstringspaces=false,
    showtabs=false,                  
    tabsize=2
}

\begin{document}

\title{Understanding Defects in Generated Codes by Language Models}


\author{\IEEEauthorblockN{Ali Mohammadi Esfahani}
\IEEEauthorblockA{\textit{Systems and Computer Engineering} \\
\textit{Carleton University}\\
Ottawa, Canada \\
alimohammadiesfahani@cmail.carleton.ca}
\and
\IEEEauthorblockN{Nafiseh Kahani}
\IEEEauthorblockA{\textit{Systems and Computer Engineering} \\
\textit{Carleton University}\\
Ottawa, Canada \\
kahani@sce.carleton.ca}
\and
\IEEEauthorblockN{Samuel A. Ajila}
\IEEEauthorblockA{\textit{Systems and Computer Engineering} \\
\textit{Carleton University}\\
Ottawa, Canada \\
samuelajila@cunet.carleton.ca}
}

\maketitle

\begin{abstract}
    
This study investigates the reliability of code generation by Large Language Models (LLMs), focusing on identifying and analyzing defects in the generated code. Despite the advanced capabilities of LLMs in automating code generation, ensuring the accuracy and functionality of the output remains a significant challenge. By using a structured defect classification method to understand their nature and origins this study categorizes and analyzes 367 identified defects from code snippets generated by LLMs, with a significant proportion being functionality and algorithm errors.  These error categories indicate key areas where LLMs frequently fail, underscoring the need for targeted improvements. To enhance the accuracy of code generation, this paper implemented five prompt engineering techniques, including Scratchpad Prompting, Program of Thoughts Prompting, Chain-of-Thought Prompting, Chain of Code Prompting, and Structured Chain-of-Thought Prompting. These techniques were applied to refine the input prompts, aiming to reduce ambiguities and improve the models' accuracy rate. The research findings suggest that precise and structured prompting significantly mitigates common defects, thereby increasing the reliability of LLM-generated code.

\end{abstract}


\lstdefinelanguage{ocl1}{
  keywords={init,Message, Pre, Post, Client,Range,Invariant,Constraints},
  keywordstyle=\color{purple}\ttfamily\bfseries,
  keywords=[2]{not, and, or, state, transition, implies, component,dbg,receipt,:,|,;},
  keywordstyle=[2]\color{black}\ttfamily\bfseries,
  sensitive=false,
  comment=[l]{//},
  morecomment=[s]{/*}{*/},
  commentstyle=\color{blue}\ttfamily,
  stringstyle=\color{red}\ttfamily,
  morestring=[b]',
  morestring=[b]"
}

\lstdefinelanguage{oc}{
  keywords={container, key,list,leaf,type},
  keywordstyle=\color{blue}\ttfamily\bfseries\footnotesize,
  keywordstyle=[2]\color{black}\ttfamily\bfseries\footnotesize,
  otherkeywords = {-,|},
  morekeywords = [3]{-},
  morekeywords = [4]{|},
  keywordstyle = [3]{\color{blue}}\ttfamily\footnotesize,
  keywordstyle = [4]{\color{blue}}\ttfamily\footnotesize,
  identifierstyle=\color{black}\ttfamily\footnotesize,
  sensitive=true,
  comment=[l]{//},
  morecomment=[s]{/*}{*/},
  commentstyle=\color{green}\ttfamily\footnotesize,
  stringstyle=\color{black}\ttfamily\footnotesize,
  morestring=[b]',
  morestring=[b]",
  alsodigit={:}
}

\lstdefinelanguage{octree}{
  keywords={rw, ro},
  keywordstyle=\color{blue}\ttfamily\bfseries\footnotesize,
  keywordstyle=[2]\color{black}\ttfamily\bfseries\footnotesize,
  otherkeywords = {-,|},
  morekeywords = [3]{-},
  morekeywords = [4]{|},
  keywordstyle = [3]{\color{blue}}\ttfamily\footnotesize,
  keywordstyle = [4]{\color{blue}}\ttfamily\footnotesize,
  identifierstyle=\color{black}\ttfamily\footnotesize,
  sensitive=true,
  comment=[l]{//},
  morecomment=[s]{/*}{*/},
  commentstyle=\color{green}\ttfamily\footnotesize,
  stringstyle=\color{black}\ttfamily\footnotesize,
  morestring=[b]',
  morestring=[b]",
  alsodigit={:}
}

\lstdefinelanguage{z3}{
	sensitive=true,
	alsoletter={\-},
	comment=[l]{;},
	keywords=[1]{
apply, assert, assert-soft, check-sat, check-sat-using, compute-interpolant,
declare-const, declare-datatypes, declare-fun, declare-map, declare-rel,
declare-sort, declare-tactic, define-sort, display, echo, eval, exit,
fixedpoint-pop, fixedpoint-push, get-assertions, get-assignment, get-info, get-
interpolant, get-model, get-option, get-proof, get-unsat-core, get-user-tactics,
get-value, help, help-tactic, labels, maximize, minimize, pop, push, query,
reset, rule, set-info, set-logic, set-option, simplify
	},
	morekeywords=[2]{
check-sat-using, declare-var, declare-rel, rule, query, set-predicate-
representation, maximize, minimize, assert-soft, assert-weighted, compute-
interpolant
	},
}
\newcommand{\oc}{\textit{OC }}
\section{Introduction}
\par
		
Pre-trained Large Language Models (LLMs) have demonstrated the capacity to manage substantial datasets and showcase exceptional performance across a diverse array of tasks~\cite{r1}. In the software development context, language models trained specifically on source code, known as Code Language Models (CLMs), have shown remarkable capabilities in automating various activities such as defect fixing, code summarization, and code generation from natural language descriptions~\cite{r7,r8,r9,r10,r11,r12,r13,r51}. This paper focuses on the latter: code generation by CLMs. Despite their impressive performance, particularly in generating syntactically correct code, these models are not without flaws. Even the best models achieve an exact match accuracy of around 20\%, and the generated code often contains defects that can hinder its practical utility, posing challenges for developers who rely on these tools.

There is a strong body of research focused on creating more effective models by improving learning algorithms, increasing model size, and incorporating large, high-quality datasets. While these efforts are foundational, achieving perfect CLMs does not seem attainable in the near future. Therefore, to effectively apply and adapt these models in the software engineering context, it is essential to understand and account for the limitations of existing CLMs to achieve successful outcomes. Although some efforts have been made in this area (e.g., \cite{r45,r46}), more research is needed.

This paper aims to analyze the defects in generated code and investigate how prompt engineering techniques can fix some of these defects. Understanding both aspects is crucial for the better application and improvement of CLMs. Identifying the types of defects not only helps to direct foundational research but also ensures that CLMs can be applied more safely, particularly in applications requiring a more conservative approach. Additionally, prompt engineering is a cost-effective technique to improve CLM results, as opposed to retraining, which is very extensive and time-consuming.  More specifically, our research address the following Research Questions (RQs):

\textbf{RQ1} \textit{What are the types of defects in the generated code, and how can they be classified based on their characteristics?} 
We selected two existing CLMs based on their availability and features, and applied these models to the well-known open-source benchmark, HumanEval. From the results, we extracted all generated code with defects. We then manually analyzed and classified these defects based on existing categorizations for source code. Additionally, we analyzed the complexity of the generated code to understand if there is a relationship between code complexity and the performance of the CLMs. \\
The results show that the majority of defects are related to Functional and Logic errors. Interestingly, there is no correlation between the correctness of the generated code and its complexity. This is surprising, as it differs from defects in developer-written code, where the probability of failure typically increases with complexity.

\textbf{RQ2} \textit{Can existing prompt engineering techniques help in fixing the problematic code?}
We studied existing prompt engineering techniques and applied them to the defective generated code identified in RQ1, focusing on their potential to produce correct code. The results, using the Exact Match (EM) metric, indicate that applying these techniques can improve outcomes by approximately 33.1\% for CodeT5+ moel and 27.8\% model for CodeGen. Among the techniques, the Structured Chain-of-Thought (SCoT) technique achieved the highest improvement.

The remainder of this paper is organized as follows: Section \ref{sec:background} provides background on language models. Section \ref{sec:relatedwork} covers related works. Section \ref{sec:eval} details our evaluation methods and discuss the results and their implications. Finally, Section \ref{sec:concl} concludes the paper.

\section{Background and Related Work}
\label{sec:background}
This section covers the core concepts on which our work is based and examines existing studies relevant to our work.
\subsection{Language Models}
Language models are designed to handle tasks involving language by estimating the probability of sequences of tokens to predict an output, typically a sequence of generated tokens \cite{r41}. CLMs are a specific type of language models tailored for programming-related activities, which are central to our research. These models are typically trained on a variety of programming languages as well as textual data \cite{r7}, \cite{r14}, \cite{r35}. CLMs are particularly effective for tasks including code summarization, code generation, program repair, and code translation. The architecture of language models predominantly draws from the integration of a self-attention layer with a feed-forward network, a concept first introduced by Vaswani et al. \cite{r42}. These components are repeated in sequences to build encoders and decoders, which are fundamental in constructing language models. In this study, the focus is on two types of language model architectures: \textit{encoder-decoder} models and \textit{decoder-only} models. \textit{Encoder-only} models are not used in this research as they lack the capability to perform generative tasks without an additional decoder \cite{r43}, and are therefore not relevant to our analysis.
\par
Encoder-decoder models are a type of model that incorporates both an encoder and a decoder. The sequence-to-sequence transformer \cite{r42} exemplifies this model type, serving as a foundation for many modern language models. In an encoder-decoder model, the encoder processes the input token sequence and transforms it into a hidden intermediary representation. This representation is then utilized by the decoder to produce the output token sequence \cite{r27}, \cite{r43}. The encoder in these models typically includes several layers, each consisting of two components: a multi-head self-attention mechanism and a fully connected feed-forward network. Similarly, the decoder comprises several layers, each equipped with three sub-layers: a masked multi-head attention mechanism, a multi-head self-attention mechanism, and a fully connected feed-forward network \cite{r42}. The multi-head self-attention mechanism applies linear transformations to the inputs, creating multiple projections of the input that are later recombined into the final output. The decoder sequentially produces the output tokens in an auto-regressive manner, choosing each subsequent token based on both the input to the model and the partially formed output sequence. In this study, we used \textit{CodeT5+} \cite{r14}-- an encoder-decoder model.
\par
Decoder-only models consist solely of a decoder, without an accompanying encoder. These models initiate with a baseline state and incrementally construct an output sequence, taking into account the tokens previously generated \cite{r27}, \cite{r44}. The functionality of the decoders in these models mirrors that found in encoder-decoder architectures, with a focus on comprehending the target language and its subtleties \cite{r27}. In this study, we used \textit{CodeGen} \cite{r35}-- a decoder-only model.

\subsection{Related work}
\label{sec:relatedwork}
Prior research has often not deeply analyzed why specific models, like CodeBERT \cite{r52} and GraphCodeBERT \cite{r53}, succeed or fail in certain scenarios. For instance, a study by Mohammadkhani et al. \cite{r46} explores how these models allocate attention to different token types during tasks such as code summarization and translation, revealing specific deficiencies when handling complex or lengthy code segments. While their work directly contributes to understanding model failures through attention-based explanations, our research goes further by providing a more comprehensive defect categorization. Unlike their study, which primarily focuses on attention allocation, this research not only identifies where models fail, but also explores why, by evaluating a broader range of defect types across multiple dimensions of code generation tasks.\\
\indent The study by Tambon et al. \cite{r45} provides a foundational analysis of bugs in code generated by LLMs, identifying ten specific bug patterns through empirical methods supported by insights from 34 industry and academic practitioners. While this work  highlights common issues such as syntax errors and misinterpretations, it covers a narrower spectrum of defect types compared to our research. In contrast, our study delves into a more exhaustive classification of defect categories, which encompasses not only syntax and semantic errors but also extends to logical and execution errors that affect the functionality of the generated code. Moreover, our research employs an approach integrating structured prompt engineering to systematically reduce these defects. 
By broadening the defect categories and introducing an intervention method that directly impacts code generation performance, our research provides a more comprehensive solution to the challenges of LLM-generated code.\\
\indent A study by Tan et al. \cite{r47} introduces the Codeflaws benchmark, a valuable contribution to the field of automated program repair through its detailed classification of 39 defect types derived from programming competition submissions. This benchmark enables a detailed evaluation of repair tools by categorizing defects based on syntactic differences between buggy and corrected programs. While the   benchmark provides a foundational approach to understanding syntactic defects, this research broadens this perspective by including semantic and logical error categories. These additional categories allow for a more comprehensive assessment of the deeper functional inaccuracies that occur in code generation by LLMs. 
This approach not only broadens the scope of defect detection but also enhances the understanding of underlying issues in code generation processes.

\section{Study Design and Results}
\label{sec:eval}
\begin{table*}[ht]
  \centering
  \caption{Defect Categories and Descriptions}
  \label{tab:bugs}
  \begin{tabular}{llp{10cm}} 
  \hline
  \textbf{Category} & \textbf{Sub-Category} & \textbf{Description}  \\
  \hline
  \textbf{Function} & \textbf{Functional error} & The intended function faces challenges in its normal implementation, deviating from the specified requirements. \\
  & \textbf{Algorithm error} & Mistakes occur in the sequence of actions employed to address a specific problem or perform a calculation, encompassing faults in the computations and faulty implementations of algorithms. \\
  \hline
  \textbf{Logic} & \textbf{Incorrect Branch} & The conditional statement within the branch is inaccurate. \\
  & \textbf{Incorrect loop} & The loop logic contains errors. \\
  & \textbf{Ignore extreme conditions} & Extreme scenarios are overlooked, lacking appropriate consideration for special cases, boundary values, and null checks. \\
  & \textbf{Redundant logic} & Additional logical statements are present.\\
  & \textbf{Conditional test error} & The logic within an ``if'' statement is incorrect. \\
  & \textbf{Logical order error} & The logical sequence or the arrangement of statements is incorrect. \\
  \hline
  \textbf{Computation} & \textbf{Incorrect operand} & The operand in the operational expression is incorrect. \\
  & \textbf{Operator error} & An incorrect operator was employed. \\
  & \textbf{Insufficient precision} & The data lacks adequate precision. \\
  \hline
  \textbf{Assignment} & \textbf{Incorrect data range or type} & The restricted data range or type is not accurate. \\
  & \textbf{Input or Output data error} & The input or output data is incorrect. \\
  \hline
  \textbf{Runtime} & \textbf{Typo} & A typographical error or mistake in code or text. \\
  & \textbf{IndexError} & Attempting to access an index that is outside the valid range of a data structure, such as an array or list. \\
  & \textbf{Type Error} & Occurs when an operation is performed on an object of an inappropriate type. \\
  & \textbf{Overflow} & When a computation result exceeds the maximum representable value, often leading to unexpected behavior or faults. \\
  & \textbf{ZeroDivisionError} &  Attempting to divide a number by zero, which is an undefined mathematical operation. \\
  \hline
  \textbf{Others} & \textbf{Others} & All the defects not in the above categories. \\
  \hline
  \end{tabular}
\end{table*}

This study aims to develop a comprehensive understanding of defects associated with code generation using LLMs by analyzing their types, underlying causes, and characteristics. To achieve this  goal, we seek to answer the following RQS:

\begin{itemize}
 \item  \textbf{RQ1} What are the types of defects in the generated code, and how can they be classified based on their characteristics? 
 
 \item  \textbf{RQ2} Can existing prompt engineering techniques help in fixing the problematic code?
 \end{itemize}

The first question focuses on analyzing and classifying the defects in the generated code, while the second investigates the effectiveness of prompt engineering techniques in addressing these issues. In the following, we describe the evaluation metrics we employ. We then explain the experimental process, present the results, and discuss their practical implications.

\subsection{Evaluation Metrics}
To evaluate the correctness of generated code by LLM, we used three evaluation metrics, including exact match accuracy (EM) \cite{r36}, pass$@$k \cite{r49}, and CodeBLEU \cite{r38}. \\
\noindent\textbf{Exact Match Accuracy (EM)}. The EM metric refers to the proportion of instances where the response from the system exactly matches the correct or expected answer. In EM, if the
model’s generated answer is an exact match with the ground truth, the EM is
1; otherwise, it is 0.

\begin{equation}
\text{EM} = 
\begin{cases} 
1 & \parbox{0.4\textwidth}{\text{if the model's answer exactly matches} \\ \text{the provided answer}} \\
0 & \text{otherwise}
\end{cases}
\end{equation}

\noindent\textbf{Pass@k}. Chen et al. \cite{r10} show that metrics like BLEU \cite{r37} fail to capture essential semantic aspects of coding, and propose adjustments to better reflect the unique features of code. Further, they highlight that match-based metrics do not consider the wide range of functionally equivalent programs that can meet the same requirements as the reference solution.
To overcome these limitations, recent studies in fields such as unsupervised code translation \cite{r48} and pseudocode-to-code translation \cite{r49} have shifted towards measuring functional correctness. In this approach, a code sample is considered correct if it successfully passes a predefined set of unit tests. This shift acknowledges that functional correctness aligns more closely with the criteria used by developers in real-world scenarios. 
In pass@k metric \cite{r49}, multiple (k) code samples are generated for each problem, and a problem is considered successfully solved if any of these samples pass all relevant unit tests. 

\begin{equation}
    \text{pass}@k := \mathbb{E}_{\text{Problems}} \left[ 1 - \frac{\binom{n-c}{k}}{\binom{n}{k}} \right]
\end{equation}

\noindent\textbf{CodeBLEU}. The BLEU metric \cite{r37}, originally designed for evaluating machine translation tasks, measures the extent of overlapping n-grams between a candidate translation and a set of reference translations. However, BLEU is inadequate for evaluating code-related tasks as it does not account for the tree structure and logical flow inherent in code. CodeBLEU \cite{r38} addresses this limitation by incorporating code-specific information into BLEU measurements. CodeBLEU integrates traditional n-gram matching with syntactic analysis using abstract syntax trees, which compares the structural alignment of code segments. It also includes semantic analysis via data-flow graphs, which checks for semantic consistency by analyzing how data moves through the code. Keyword matching evaluates the use of language-specific keywords that are critical for code’s functionality. The CodeBLEU metric is defined as follows:

\begin{align}
\text{CodeBLEU} &= \alpha \cdot \text{BLEU} + \beta \cdot \text{SM} \notag \\
&\quad + \gamma \cdot \text{DM} + \delta \cdot \text{KM}
\end{align}
For this study, initial values for  \( \alpha, \beta, \gamma, \sigma \) are set heuristically, reflecting a balanced contribution from each component, with plans for empirical optimization in future work.
\subsection{Dataset}
To assess the performance of LLMs in code generation tasks, we utilized a collection of 164 handwritten programming problems known as the HumanEval \cite{r10, r26} dataset. Each problem within this dataset comprises a function signature, a docstring, the function's body, and multiple unit tests, with an average of 7.7 tests per problem. These programming tasks are designed to evaluate language understanding, logical reasoning, algorithmic thinking, and basic mathematical skills. 


\begin{figure*}[ht]
\centering

\begin{subfigure}[b]{0.48\textwidth}
\begin{lstlisting}
def generate_integers(a, b):
    """
    Given two positive integers a and b, return the even digits between a and b, in ascending order.
    """
    
    if a > b:
    
        a = b
        b = a
        
    even_numbers = []
    while a!= 0:
        even_numbers.append(a % 10)
        a //= 10
    return even_numbers[::-1]

\end{lstlisting}
\caption{ Incorrect Loop Error}
\label{lst:loop}
\end{subfigure}
\hfill
\begin{subfigure}[b]{0.48\textwidth}
\begin{lstlisting}
def eat(number, need, remaining):
    """
    You're a hungry rabbit who has already eaten some carrots but needs more to finish your daily meals. Return an array showing the total carrots eaten and those remaining. If insufficient, eat all remaining carrots but remain hungry.
    """
    if number == 0:
        return [0, 0]
    if need == 0:
        return [0, 0]
    if remaining == 0:
        return [0, 0]
    if number > need:
        return [number - need, need]
    else:
        return [0, 0]

\end{lstlisting}
\caption{Algorithm Error}
\label{lst:Algorithm}
\end{subfigure}

\vspace{0.5cm} 

\begin{subfigure}[b]{0.48\textwidth}
\begin{lstlisting}
def sum_of_two_digits(arr, k):
    """
    Given a non-empty array of integers arr and an integer k, return the sum of the elements with at most two digits from the first k elements of arr.
    """
    if k > len(arr):
        return 0
    else:
        return sum(arr[k:]) + sum(arr[:k])

\end{lstlisting}
\caption{Conditional Test Error}
\label{lst:Conditional}
\end{subfigure}
\hfill
\begin{subfigure}[b]{0.48\textwidth}
\begin{lstlisting}
def odd_digits(n):
    """
    Given a positive integer n, return the product of the odd digits. Return 0 if all digits are even. For example: digits(1) == 1 digits(4) == 0 digits(235) == 15
    """
    
    if n % 2 == 0:
        return n
    else:
        return 0
\end{lstlisting}
\caption{Functionality Error}
\label{lst:Fuctionality}
\end{subfigure}
\caption{Examples of generated  defective codes by LLMs}
\label{fig:buggycode}
\end{figure*}



    \par

\subsection{Defect Classification}
To provide a better understanding of defects in generated code, we have classified them into six categories and nineteen subcategories. Our classification builds upon the work of Ni et al. \cite{r17}, who examined key aspects of defect resolution and the origins of these defects using the Orthogonal Defect Classification (ODC) method \cite{r18}. This approach offers a comprehensive framework for analyzing defects, with particular emphasis on localization, elimination, cause analysis, and prevention. The ODC method classifies defects into eight distinct attributes: Activity, Triggers, Impact, Target, Type, Qualifier, Source, and Age, focusing specifically on attributes related to source code. Ni et al.'s study \cite{r17} investigated the correlation between defects fixes and the causes of defects, using the ODC method to categorize causes based on defect type attributes related to the source code.

To have a more refined categorization, we have also integrated the IEEE classification method into our analysis. The IEEE standard for classifying software anomalies \cite{r19} provides a detailed framework for categorizing a wide spectrum of software issues, outlining a systematic approach to managing software anomalies throughout the software lifecycle. This classification is executed through four sequential steps, interspersed with three administrative activities, ensuring a thorough characterization of software defects.  The table \ref{tab:bugs} outlines our classification, which includes six categories: \textit{Function, Logic, Computation, Assignment, Runtime,} and \textit{Others}. 
\begin{itemize}
    \item 
\textbf{Function category} refers to defects include logical inconsistencies, unexpected behaviors, or outcomes that do not align with the desired code behavior, indicating a gap between what is expected and what is produced. We have identified two subcategories including  \textit{functional error} and \textit{algorithm error}. Functionality errors represent discrepancies between the generated code, and the specified requirements. Algorithm errors point towards inaccuracies in the algorithmic steps employed to solve a particular problem or calculation. Such discrepancies result from faults in calculations, faulty implementations of algorithms, or a misinterpretation of the underlying computational logic.
   \item \textbf{Logic category} refers to defects in the logical structure of the code, such as incorrect branch and loop, ignore extreme conditions, redundant logic, conditional test error, and logical order error. The details of these defects can be found in Table \ref{tab:bugs}. 
   \item \textbf{Computation category} refers to defects  in the mathematical and computational operations within the code, which includes incorrect operand, operator error, and insufficient precision.
   \item \textbf{Assignment category} refers to defects related to data assignment, which includes incorrect data range or type, and input or output data error.
   \item \textbf{Runtime category} refers to defects that occur during the execution of the code, which includes typo, index error, type error, overflow, and zeroDivisionError. 
   \item \textbf{Others category} refers to defects not covered by the above categories. 
   \end{itemize}

\subsection{Model Selection}

A comprehensive review by Hou et al. \cite{r27} examined the use of LLMs in software engineering, identifying over 50 models applicable to various tasks within this field.  Given the impracticality of evaluating all available LLMs due to computational constraints, we have developed specific criteria to select the LLMs for our study addressing RQ1.\\
\textbf{Open Source}. The study focused on models that are open-source to facilitate replicability. This criterion excludes models such as GPT-4\cite{r28}, AlphaCode\cite{r29},  PaLM-Coder\cite{r30}, and Codex\cite{r10} .\\
\textbf{Pre-Trained on Code}.
This study aims to understand defects in LLM-generated code, so it is necessary to use models that have been specifically pre-trained on large code datasets. As a result, we are not considering models trained on general natural language datasets. For instance, models such as T5\cite{r31}, GPT-Neo\cite{r32}, and GPT-J\cite{r33} are excluded based on this criterion.\\
\textbf{Old Models}.
Models that have been outdated by more recent versions were excluded from our study. For example, we have not included CodeT5\cite{r34} because it has been replaced by its updated version, CodeT5+\cite{r14}.
\par
Using the above criteria, we limited our selection to two LLMs: \textit{CodeT5+} \cite{r14}, an encoder-decoder model with 770 million parameters, and \textit{CodeGen Multi} \cite{r35}, a decoder-only model with 350 million parameters.

\begin{table*}[ht]
  \centering
  \caption{Distribution of Defects Identified in Generated Codes by LLMs}
  \label{tab:bug_distribution}
  \begin{tabular}{llcc}
    \toprule
    \textbf{Defect Category} & \textbf{Sub-Category} & \textbf{Number of Occurrences} & \textbf{Percentage of Occurrences}\\
    \midrule
    Function & Functionality Error & 116 & 31.6\% \\
    & Algorithm Error & 108 & 29.3\% \\
    Logic & Conditional Test Error & 89 & 24.1\% \\
    & Incorrect Loop & 17 & 4.60\%\\
    & Incorrect Branch & 6 & 1.74\% \\
    & Ignore Extreme Conditions & 2 & 0.57\% \\
    & Redundant Logic & 2 & 0.57\% \\
    Computation & Incorrect Operand & 2 & 0.57\% \\
    Assignment & Incorrect Data Range or Type & 4 & 1.15\% \\
    Runtime & Typo & 4 & 1.15\%\\
    & IndexError & 4 & 1.15\% \\
    & Type Error & 2 & 0.57\% \\
    & Overflow & 2 & 0.57\% \\
    & ZeroDivisionError & 2 & 0.57\% \\
    Other & Wrong Comment & 7 & 1.90\% \\
    \bottomrule
  \end{tabular}
\end{table*}

\subsection{RQ1 What are the types of defects in the generated code, and how can they be classified based on their characteristics?}

\subsubsection{Overall Results} To answer this question, we presented each prompt from the benchmark to the selected models and executed each generated code snippet by the models to test their functionality against the benchmark's test cases. For the CodeT5+ model, our results indicated that 26 instances (15.85\%) successfully produced accurate and functional code—passed all test cases. However, 34 instances (20.73\%) either failed to generate any code, produced incomplete snippets, or simply returned the given prompt without a code snippet. Furthermore, 104 instances (64.41\%) generated code but contained one or more defects.

Similarly, for the CodeGen model, out of 164 instances, only 11 instances (6.67\%) successfully produced accurate and functional code, passing all test cases. Additionally, 32 instances (19.51\%) failed to generate any code or returned incomplete snippets, while 121 instances (73.78\%) generated code that contained one or more defects. These findings highlight the challenges LLMs face in producing error-free and complete code under automated conditions.

Exact Match (EM), pass@k (with K=5), and CodeBLEU are used to assess the correctness of the code generation models. The EM metric for CodeT5+ showed that 15.85\% of the code generated matched the ground truth, indicating an understanding of the tasks. For the CodeGen model, this metric significantly dropped to 6.67\%, emphasizing the model's challenges in achieving syntactic and semantic precision. 

The pass@k metric, which evaluates the likelihood of achieving at least one correct solution within five attempts for each problem, showed an effectiveness of 20.7\% for CodeT5+ but was much lower for CodeGen, standing at 8.42\%. This metric is particularly valuable as it provides a more lenient and realistic measure of a model's capability in practical applications, compared to the more stringent EM metric.

For the CodeBLEU metric, which combines syntactic and semantic evaluation with traditional n-gram overlap, the average score for CodeT5+ was 29.5\%, indicating moderate success in generating code that, while not always perfect, maintains high degrees of coherence and functionality. The score for CodeGen was lower, at 20\%, further highlighting the model's issues with generating functionally and syntactically coherent code. This metric provides insight into the practical utility of generated code in real coding environments.

The evaluation metrics employed to assess the code generation performance of the models are summarized in Table \ref{tab:metrics}. This table provides a clear understanding of the accuracy of the different models used in this study.

\begin{table}[h]
\centering
\caption{Evaluation Metrics for Code Generation}
\label{tab:metrics}
\begin{tabular}{lcc}
\hline
\textbf{Metric} & \textbf{CodeT5+ Value (\%)} & \textbf{CodeGen Value (\%)} \\
\hline
Exact Match (EM) & 15.85 & 6.67 \\
pass@k (k=5) & 20.7 & 8.42 \\
CodeBLEU & 29.5 & 20 \\
\hline
\end{tabular}
\end{table}

\subsubsection{Classification of the Defects}
\label{subsec:cof}
We conducted a detailed manual inspection of the code snippets generated by the LLMs to identify potential defects. The manual review of code generated by LLMs revealed 367 defects, categorized in Table \ref{tab:bug_distribution}, which provides a detailed breakdown of various defect categories and their respective frequencies. In the following, we discuss the most frequent defects that appeared in the generated code based on their category.

\textbf{Function Category}. \textit{Algorithm  and functionality errors} were the main defects, appearing in 108 (29.3\%) and 116 (31.6\%) instances, respectively. These defects highlight  challenges in the model's capacity to understand and correctly execute the intended functionality. For instance, given task to the model is (Figure \ref{lst:Fuctionality}): ``Given a positive integer ``n'', return the product of the odd digits. Return 0 if all digits are even''. The generated code checks whether the given integer ``n'' is even and returns ``n'' if it is. However, the task requires finding the product of the odd digits. Algorithm errors, on the other hand, account for 108 instances, indicating a distinct set of challenges related to the computational aspects of the generated code. Algorithm errors highlight a crucial aspect where the model's understanding of the specified computational procedures fails, resulting in deviations from the intended problem-solving approach. For instance, the given task is (Figure \ref{lst:Algorithm}): ``You're a hungry rabbit, and you already have eaten a certain number of carrots, but now you need to eat more carrots to complete the day's meals. you should return an array of [total number of eaten carrots after your meals, the number of carrots left after your meals] if there are not enough remaining carrots, you will eat all remaining carrots, but will still be hungry.''. The   generated code  does not correctly calculate the total number of eaten carrots and the carrots left after the meals. Instead, it returns in various conditions without considering the actual calculations based on the input parameters. 

\textbf{Logic Category}. \textit{Conditional test error} occurs  in 89 (24.1\%) instances. For example, the task is (Figure \ref{lst:Conditional}): ``given a non-empty array of integers arr and an integer k, return the sum of the elements with at most two digits from the first k elements of arr''. The generated code instead  summed all elements after the first k elements. 
 \textit{Incorrect loop error}  where the loop logic is not correct, appeared in 17 instances (24.1\%). Example of this defect (Figure \ref{lst:loop}), where the task is: ``Given two positive integers a and b, return the even digits between a and b, in ascending order''. The generated code used a while loop that continued until ``a'' becomes zero. The generated code  did not consider the range of numbers between a and b.

\subsubsection{Complexity of the Defective Generated Code} In addition to classifying the defects, we examined the complexity of the expected codes for all samples to determine if the defects in the generated code by the LLM models are related to the code complexity. In our study, we used \textit{cyclomatic complexity} to assess the complexity of the generated code. Cyclomatic complexity measures the number of linearly independent paths through a program's source code \cite{r40}.

Given the nature of our data, which consists of a binary outcome (1 for success and 0 for failure) and continuous data for cyclomatic complexity, we employed logistic regression \cite{r50} to perform correlation analysis. Logistic regression is well-suited for scenarios where the dependent variable is categorical, allowing us to model the probability of success as a function of cyclomatic complexity. The results of the logistic regression indicated a Pseudo R-squared value of 0.005523, suggesting that the model explains a very small portion of the variance in success rates. Additionally, the logistic regression model yielded a non-significant p-value (0.418) for the coefficient of cyclomatic complexity, indicating that increases in complexity do not significantly decrease the likelihood of successful code generation by LLMs.
 
Furthermore, the correlation analysis between cyclomatic complexity and success rate yielded a coefficient of -0.063, indicating a very weak negative relationship. This weak correlation supports the logistic regression results, suggesting that complexity alone may not be a decisive factor in predicting code generation success. This suggests the need for new research in this area to understand the root causes of model failure. It further motivates our next research question, where we investigate how do prompt engineering techniques impact the quality of the code generated by LLMs.

\subsection{RQ2 Can existing prompt engineering techniques help in fixing the problematic code?}

\begin{figure*}[ht]
\centering

\begin{subfigure}[b]{0.48\textwidth}
\begin{lstlisting}
Q: Given a positive integer n, return the product of the odd digits. Return 0 if all digits are even.
Example: n = 235
1. Initialize product as 1 and an empty list for odd digits.
2. Split the number into its digits: 2, 3, 5
3. Check each digit:
   - 2 is even, skip.
   - 3 is odd, add to list and multiply product: product = 3
   - 5 is odd, add to list and multiply product: product = 3 * 5 = 15
4. If no odd digits, return 0.
Answer: 15
\end{lstlisting}
\caption{Scratchpad Prompting}
\label{lst:Scratchpad}
\end{subfigure}
\hfill
\begin{subfigure}[b]{0.48\textwidth}
\begin{lstlisting}
Q: Given a positive integer n, return the product of the odd digits. Return 0 if all digits are even.
Example: n = 235
```python
def product_of_odd_digits(n):
    digits = [int(d) for d in str(n)]
    odd_digits = [d for d in digits if d % 2 != 0]
    if not odd_digits:
        return 0
    product = 1
    for d in odd_digits:
        product *= d
    return product

result = product_of_odd_digits(235)
print(result)  # Output: 15
\end{lstlisting}
\caption{Chain of Code (CoC) Prompting}
\label{lst:CoC}
\end{subfigure}

\vspace{0.5cm} 

\begin{subfigure}[b]{0.48\textwidth}
\begin{lstlisting}
Q: Given a positive integer n, return the product of the odd digits. Return 0 if all digits are even.
Example: n = 235
1. Split the number into its digits: 2, 3, 5
2. Identify the odd digits: 3, 5
3. Calculate the product of the odd digits: 3 * 5 = 15
Answer: 15
\end{lstlisting}
\caption{Chain-of-Thought (CoT) Prompting}
\label{lst:CoT}
\end{subfigure}
\hfill
\begin{subfigure}[b]{0.48\textwidth}
\begin{lstlisting}
Q: Given a positive integer n, return the product of the odd digits. Return 0 if all digits are even.
Example: n = 235
1. Convert the number to a list of digits: [2, 3, 5]
2. Filter out the even digits: [3, 5]
3. Multiply the odd digits: 3 * 5 = 15
4. Return the product or 0 if no odd digits.
Answer: 15
\end{lstlisting}
\caption{Program of Thoughts (PoT) Prompting}
\label{lst:PoT}
\end{subfigure}

\begin{subfigure}[b]{\textwidth}
\centering
\begin{lstlisting}
Q: Given a positive integer n, return the product of the odd digits. Return 0 if all digits are even.
Example: n = 235

1. Initialize product as 1 and a flag foundOdd as False.
2. Convert n to a string and iterate over each character to process each digit:
   - Start loop:
     - `2` is even, skip.
     - `3` is odd, update product: product = 3, set foundOdd = True.
     - `5` is odd, update product: product = 3 * 5 = 15.
   - End loop.
3. Final check:
   - If foundOdd is False (no odd digits found), return 0.
   - Otherwise, return the product.

Output: 15 (product of odd digits 3 and 5)
\end{lstlisting}
\caption{Structured Chain-of-Thought (SCoT) Prompting}
\label{lst:SCoT}
\end{subfigure}

\caption{Examples of code generation for each prompt engineering technique}
\label{fig:PromptEngineeringTechniques}
\end{figure*}

As discussed in RQ1, a significant portion, namely 34\%, of the defects identified pertained to functionality errors. 
In addressing the issues discussed in RQl, one key challenge with LLMs is the potential for misunderstandings between humans and LLMs. Questions that seem clear to humans may be misinterpreted by LLMs. To improve the reliability of these models, we used five prompt engineering techniques commonly applied to code generation by LLMs: Scratchpad Prompting \cite{r20}, Program of Thoughts (PoT) Prompting \cite{r21}, Chain-of-Thought (CoT) Prompting \cite{r22}, Chain of Code (CoC) Prompting \cite{r23}, and Structured Chain-of-Thought (SCoT) Prompting \cite{r39}. These techniques enhance the effectiveness of our dataset prompts. Each prompt in the dataset was carefully engineered based on the specific technique used. In the following, we describe these prompt engineering techniques and demonstrate their application to the odd digits example discussed in Section  \ref {subsec:cof} (also shown in Figure~\ref{lst:Fuctionality}). We then present the results of their application.

\textbf{Scratchpad Prompting}. 
Scratchpad \cite{r20} provides a mechanism for the model to generate and record intermediate steps or thoughts, which helps in breaking down complex tasks into simpler, manageable parts.   This method logs each computational step, ensuring clarity and accuracy.
Tuning the scratchpad model is crucial to ensure that it captures and processes these intermediate steps effectively. The tuning involves structuring the training data to reflect the detailed breakdown of tasks into smaller subtasks, mimicking a step-by-step human problem-solving approach. Each training example consists of a sequence of operations with their expected intermediate outputs, teaching the model to progressively build towards the final answer. In tasks requiring multi-step calculations like integer addition, training examples explicitly detail each arithmetic step and the intermediate sums. This method  
improves out-of-distribution generalization by exposing the model to a variety of computational pathways during training.

In the context of the running example that asks the model to calculate the  product of the odd digits in a given number (e.g., 235), the process begins by initializing the product to 1 (input: none, output: product = 1) and creating an empty list for odd digits (input: none, output: empty list). The model then processes the integer 235 by splitting it into its digits (input: 235, output: [2, 3, 5]). Each digit is evaluated individually: the digit 2 is even and is skipped (input: 2, output: none), while the digits 3 and 5 are identified as odd. The product is updated sequentially with each odd digit (input: 3, output: product = 3; input: 5, output: product = 15). The final product of 15 is obtained after multiplying the odd digits (input: [3, 5], output: product =15). These steps can be seen in Figure  \ref{lst:Scratchpad}. 
\par
\textbf{Program of Thoughts (PoT) Prompting.} 
 PoT  ~\cite{r21} leverages LLMs to facilitate step-by-step reasoning and computation for complex numerical reasoning tasks. Unlike traditional methods that combine reasoning and computation within the model, PoT distinctly separates these processes. It allows the  model to focus on generating a programmatic description of the reasoning process in a high-level programming language, such as Python. The actual computations are then performed by an external interpreter. This separation enhances the model's performance by reducing computational faults common in language models, particularly with complex or iterative calculations. Also, PoT supports more structured reasoning by enabling detailed programming instructions, which can include loops, conditional statements, and function calls, leading to more accurate and interpretable solutions.

In the context of the example, PoT first defines a Python function, product-of-odd-digits, to compute the product of odd digits from the input integer (input: function definition, output: function structure). The function processes the number 235 by converting it into a list of its digits (input: 235, output: [2, 3, 5]). A list comprehension filters out even digits, leaving only the odd digits (input: [2, 3, 5], output: [3, 5]). It initializes a product variable at 1 and iterates over the odd digits, multiplying them together (input: [3, 5], output: product = 15). If no odd digits were present, the function would return 0 (input: empty list, output: 0). But, in this case, it computes the product to be 15 (input: execute function with 235, output: 15). The function is executed to provide the result (Figure \ref{lst:PoT}).

\textbf{Chain-of-Thought (CoT) Prompting.} This technique enhances the reasoning capabilities of language models by guiding them to articulate a logical sequence of thoughts leading to a solution \cite{r22}. This  encourages models to simulate a human-like problem-solving approach where each step in reasoning is explicitly stated before arriving at the final answer. 
By employing CoT prompting, models are better equipped to tackle a variety of complex reasoning tasks, such as arithmetic, commonsense, and symbolic reasoning.  
CoT does not require extensive retraining with large datasets but instead leverages few-shot learning, where the model learns from a small number of examples to generalize its reasoning capabilities to new tasks. The transparency provided by CoT also offers a valuable window into the model's cognitive process, allowing for easier debugging and understanding of how models arrive at certain conclusions.

In the context of the running example, CoT method begins by splitting the number 235 into its individual digits (input: 235, output: [2, 3, 5]). It identifies which digits are odd (input: [2, 3, 5], output: [3, 5]) and then calculates the product of these odd digits (input: [3, 5], output: 15). Each step is carefully detailed, from the initial input of the integer to the intermediate identification of odd digits and the final computation of their product. The final answer of 15 (input: calculate product, output: 15) reflects a clear sequence of thought and computation (Figure \ref{lst:CoT}).

\textbf{Chain of Code (CoC) Prompting.} 
CoC  \cite{r23}  improves upon the traditional CoT \cite{r22} by integrating executable code snippets, which allows language models not only to generate code but also to execute it. 
The technique works by encouraging language models to format solutions as flexible pseudocode. When executed, this pseudocode is processed directly by a code interpreter for computable tasks, or simulated by the language model itself for tasks that involve semantic understanding or are not directly executable. This dual capability  expands the types of problems the model can solve, from straightforward computational tasks to complex reasoning problems involving semantic analysis.

In the context of the example, CoC starts with the integer 235, which the model translates into executable code that first parses this integer into a list of its individual digits (input: 235, output: [2, 3, 5]). It then filters this list to retain only the odd digits (input: [2, 3, 5], output: [3, 5]). Following this, the model computes the product of the retained odd digits (input: [3, 5], output: 15). The entire sequence—extraction, filtering, and multiplication—is dynamically generated as a coherent block of code and executed to confirm the solution (input: generated code execution, output: 15). These steps can be seen in Figure \ref{lst:CoC}.

\begin{table*}[ht]
\centering
\caption{Performance Comparison of Different Prompt Engineering Techniques on CodeT5+ and CodeGen Models}
\label{tab:comparison}
\begin{tabular}{@{}lcccc@{}}
\toprule
 & \multicolumn{2}{c}{\textbf{CodeT5+}} & \multicolumn{2}{c}{\textbf{CodeGen}} \\ 
\cmidrule(lr){2-3} \cmidrule(lr){4-5}
\textbf{Method} & \textbf{EM} & \textbf{Improvement} & \textbf{EM} & \textbf{Improvement} \\ 
\midrule
Scratchpad Prompting & 28.6\% & +12.75\% & 21.4\% & +14.73\% \\
Program of Thoughts (PoT) Prompting & 31.2\% & +15.35\% & 22.6\% & +15.93\% \\
Chain-of-Thought (CoT) Prompting & 27.9\% & +12.05\% & 17.6\% & +10.93\% \\
Chain of Code (CoC) Prompting & 32.3\% & +16.45\% & 26.2\% & +19.53\% \\
Structured Chain-of- Thought (SCoT) Prompting & 33.1\% & +17.25\% & 27.8\% & +21.13\% \\
\bottomrule
\end{tabular}
\end{table*}

\textbf{Structured Chain-of-Thought (SCoT).} 
This technique \cite{r39} refines the concept of CoT by integrating structured programming paradigms into the reasoning process of language models. It asks the model to first conceptualize the problem-solving steps using common programming structures—sequence, branch, and loop—before generating the actual code. This method leverages the structured nature of programming to enhance the model’s ability to generate syntactically correct and logically coherent code.

SCoT guides the model to think in terms of control flows and data structures, reflecting a more accurate simulation of a programmer's thought process when faced with a coding task. 

In the context of the example, SCoT begins by initializing a product variable to 1 and a boolean flag foundOdd to False, preparing for the processing of each digit within the integer n (input: None, output: product = 1, foundOdd = False). The integer 
n is then converted into a string to allow iteration over each character, representing the digits (input: 235, output: `2', `3', `5').
As the iteration proceeds, each digit is evaluated to determine its parity. Even digits are bypassed, while odd digits trigger an update to the product variable, which is multiplied by the digit value, and the foundOdd flag is set to True (input: `3', output: product = 3; input: `5', output: product = 15). This sequence of checks and updates—initialization, digit evaluation, and conditional updating—ensures that each digit is processed. Following the completion of the loop, a final evaluation checks the foundOdd flag. If no odd digits were found (foundOdd = False), the function returns 0, indicating the absence of odd digits (input: foundOdd = False, output: 0). Otherwise, the accumulated product of the odd digits is returned (input: foundOdd = True, output: 15). These steps can be seen in Figure \ref{lst:SCoT}. \\
\indent The distinctions among the prompt engineering techniques that used in this study focus on how they structure the reasoning process of language models to enhance their performance across various tasks. Scratchpad Prompting enables models to keep a record of intermediate calculations, which is crucial for tasks with complex, multi-step processes. This method is uniquely different from PoT, which structures the reasoning into a program-like format, allowing the model to plan before executing tasks, thereby reducing the load during computation.
CoT, in contrast, emphasizes enhancing transparency by encouraging the model to express its reasoning in a detailed, stepwise manner in natural language. This approach is beneficial for educational purposes where understanding the model's reasoning is crucial, differing significantly from CoC, which integrates executable code snippets directly into the model’s reasoning. CoC ensures that each reasoning step is not only logical but also executable, blending logical reasoning directly with code generation.
SCoT further refines CoT's approach by structuring the reasoning process into typical programming constructs such as loops and conditions, making the output more structured like actual programming code. 
Each technique offers unique advantages depending on the specific needs of the task, whether it is managing complex computations, ensuring clear logical progression, or producing directly executable code outputs.

\subsubsection*{Results}

We applied each of the discussed prompt engineering techniques to all defective samples within the benchmark. This systematic application allowed us to comprehensively evaluate the effectiveness of each technique in enhancing the performance of generated code  by CodeT5+ and CodeGen models. The results  are shown in Table \ref{tab:comparison}. These improvements were measured using the EM metric, which gauges the accuracy of the code generated by the models.
The data reveals that each technique  enhances the model's code generation capabilities by refining the prompts to reduce ambiguities and improve logical flow in the generated code.

The SCoT prompting technique yielded the highest improvement, with a 17.25\% increase for CodeT5+ and a 21.13\% increase for CodeGen. This technique effectively guides the language model through a structured reasoning process, enhancing the model’s ability to handle complex coding tasks by breaking them down into manageable, logically connected steps.
Conversely, the CoT prompting technique, while still beneficial, provided the least improvement among the techniques we tested. This method improved EM by 12.05\% for CodeT5+ and 10.93\% for CodeGen. Although it offers clear advantages, its relatively lower performance suggests that it may be more suited to tasks that require straightforward sequential reasoning rather than complex algorithmic processes.

The PoT prompting and Scratchpad prompting techniques also showed notable improvements. PoT resulted in a 15.35\% improvement for CodeT5+ and a 15.93\% improvement for CodeGen. Scratchpad prompting improved EM by 12.75\% for CodeT5+ and 14.73\% for CodeGen.
Lastly, the CoC prompting technique provided  improvements, with a 16.45\% increase for CodeT5+, and a 19.53\% increase for CodeGen.

\vspace{-.2cm}
\section{Conclusion}
\label{sec:concl}

Our analysis of code generation by Code Language Models (CLMs) revealed challenges. This paper categorizes and analyzes 367 defects identified primarily as functionality and algorithmic errors. These errors indicate areas where LLMs frequently fail and the need for improvement. The paper also demonstrated that the prompt engineering techniques can improve code accuracy and reliability. On average, the techniques improved the Exact Match (EM) accuracy by approximately 15.44\% for CodeT5+ and 16.45\% for CodeGen, with the SCoT Prompting showing the most significant gains at 17.25\% and 21.13\% respectively. This suggests that future research should focus on developing advanced prompt engineering strategies and assessing their impact across diverse coding environments. Future studies could expand the model comparison to include potentially more capable LLMs such as CodeLlama, GEMMA, and Llama3. Additionally, a detailed examination of how each prompt engineering technique affects specific categories of defects could be included. Such analysis would enable a more targeted approach in applying these techniques based on the nature of the defect, potentially leading to more refined improvements in code generation accuracy and reliability.

\bibliographystyle{IEEEtran}
\bibliography{refrences}

\end{document}